# Assessment of saddle-point-mass predictions for astrophysical applications

Aleksandra Kelić and Karl-Heinz Schmidt

*GSI, Planckstr. 1, D-64291 Darmstadt, Germany*

**Abstract.** Using available experimental data on fission barriers and ground-state masses, a detailed study on the predictions of different models concerning the isospin dependence of saddle-point masses is performed. Evidence is found that several macroscopic models yield unrealistic saddle-point masses for very neutron-rich nuclei, which are relevant for the r-process nucleosynthesis.

## INTRODUCTION

In order to have a full understanding of the r-process nucleosynthesis it is indispensable to have the proper knowledge of the fission process. In the r-process, fission has the decisive influence on the termination of the r-process as well as on the yields of transuranium elements and, consequently, on the determination of the age of the Galaxy and the Universe [1]. In cases where high neutron densities exist over long periods, fission will also influence the abundances of nuclei in the region $A \sim 90$ and 130 due to the fission cycling [2,3].

First studies on the role of fission in the r-process have been started forty years ago [2]. In the meanwhile, extensive investigations on the beta-delayed and neutron-induced fission have been performed; see e.g. [4,3,5,6]. Recently, first studies on the role of neutrino-induced fission in the r-process have also been done [7,8]. One of the common conclusions from all this work is that the influence of fission on the r-process is very sensitive to the fission-barrier heights of heavy r-process nuclei with $A > 190$ and $Z > 84$, since they determine the calculated fission probabilities of these nuclei. Unfortunately, experimental information on the height of the fission barrier is only available for nuclei in a rather limited region of the chart of the nuclides. Therefore, for heavy r-process nuclei one has to rely on theoretically calculated barriers. Due to the limited number of available experimental barriers, in any theoretical model the constraint on the parameters defining the dependence of the fission barrier on neutron excess is rather weak. This leads to large uncertainties in estimating the height of the fission barriers of heavy nuclei involved in the r-process. For example, it was shown in Ref. [5] that predictions on the beta-delayed fission probabilities for nuclei in the region $A \sim 250 - 290$ and $Z \sim 92 - 98$ can vary between 0% and 100 % depending on the mass model used (see e.g. Table 2 of Ref. [5]), thus strongly influencing the r-process termination point. Moreover, the uncertainties within the nuclear models used to calculate the fission barriers can have important consequences on the r-process. Meyer *et al.* have shown that a change of 1 MeV in the fission-barrier height can have a strong influence on the production of the progenitors ($A \sim 250$) of the actinide cosmochronometers and, thus, on the nuclear cosmochronological age of the Galaxy [9].

Recently, important progress has been made in developing full microscopic approaches to nuclear masses (see e.g. [10]). Nevertheless, due to the complexity of the problem, this type of calculations is difficult to apply to heavy nuclei, where one is still to deal with semi-empirical models. Often used models are of the macroscopic-microscopic type, where the macroscopic contribution to the masses is based either on some liquid-drop, droplet or Thomas-Fermi model, while microscopic corrections are calculated separately, mostly using the Strutinsky method [11]. The free parameters of these models are fixed using the nuclear ground-state properties and, in some cases, the height of fission barriers when available. Some examples of such calculations are shown in Fig. 1 (upper part), where the fission-barrier heights given by the results of the Howard-Möller fission-barrier calculations [12], the finite-range liquid drop model (FRLDM) [13], the Thomas-Fermi model (TF) [14], and the extended Thomas-Fermi model with Strutinsky integral (ETFSI) [15] are plotted as a function of the mass number for several uranium isotopes ($A$ = 200-305). In case of the FRLDM and the TF model, the calculated ground-state shell corrections of Ref. [16] were added as done in Ref. [18]. In cases where the fission barriers were measured, the experimental values are also shown.

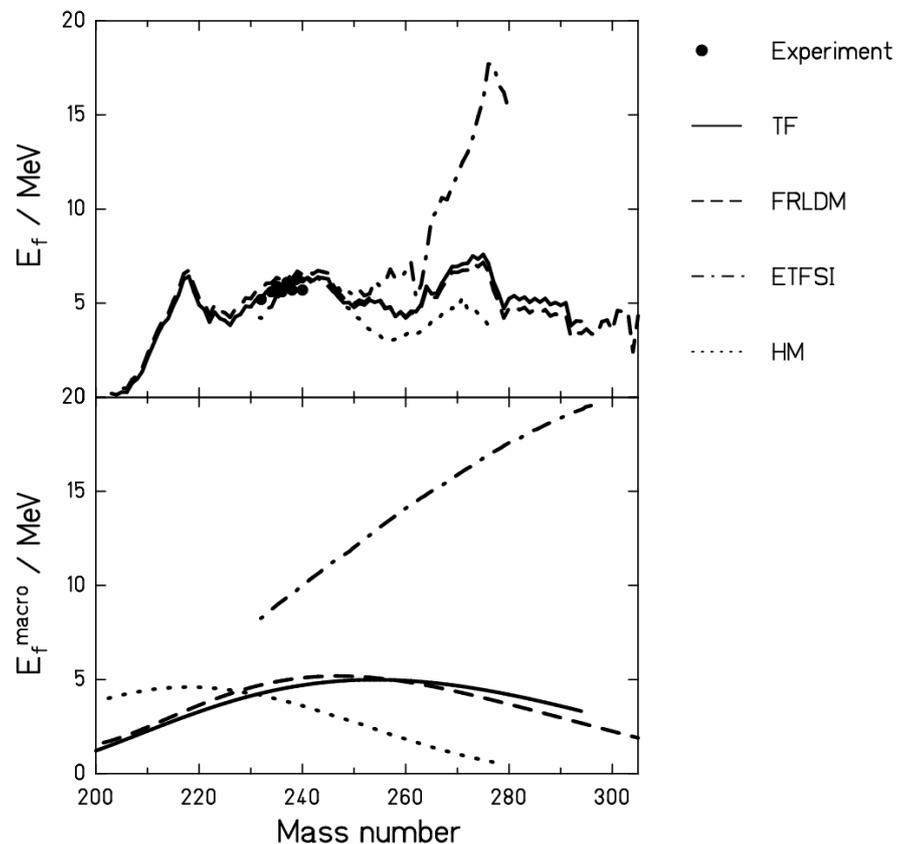

**FIGURE 1.** Full macroscopic-microscopic (upper part) and macroscopic part (lower part) of the fission barrier calculated for different uranium isotopes using: the extended Thomas-Fermi model + Strutinsky integral [15] (dashed-dotted line), the finite-range liquid-drop model [13] (dashed line), the Thomas-Fermi model [14] (full line), and the Howard-Möller tables [12] (point-line). In case of FRLDM and TF the ground-state shell corrections were taken from Ref. [16]. The macroscopic part of the Howard-Möller results is based on the droplet model [17]. The existing experimental data shown in the upper part of the figure are taken from the compilation in Ref. [19].

From the figure it is clear that as soon as one enters the experimentally unexplored region there is a severe divergence between the different model predictions. Of course, these differences can be caused by both – macroscopic and microscopic – parts of the models, but in case of nuclei with non-vanishing macroscopic fission barriers, the topographic theorem of Myers and Swiatecki [18] requires that the macroscopic part of the saddle-point masses should be very close to the measured saddle-point masses, and, consequently, the observed differences are most likely to be attributed to failures in the macroscopic model. The macroscopic fission barriers, calculated with the above-mentioned models[1] for the same uranium isotopes, are shown in the lower part of Fig. 1.

Therefore, in this paper, we consider the macroscopic part of the above-mentioned models and study the behaviour of the macroscopic contribution to the fission barriers when extrapolating to very neutron-rich nuclei. This study is based on the approach of Dahlinger *et al.* [19], where the predictions of the theoretical models are examined by means of a detailed analysis of the isotopic trends of ground-state and saddle-point masses. It is not our intention to develop a new model for calculating fission barriers or to suggest possible improvements in already existing model. The goal of this paper is to test the existing models and to suggest those which are the most reliable to be used in astrophysical applications.

## METHOD

Usually, when one tests the predicted fission barriers of a theoretical model, one compares the heights of experimentally determined and calculated fission barriers. In doing so, one is obliged to use the theoretically calculated ground-state shell corrections, which can introduce an additional important uncertainty in the model predictions. To avoid this problem, we have decided to compare the measured and the model-calculated saddle-point masses, as was already suggested in [19]. As noted above, we will only consider the macroscopic saddle-point masses given by the following models:

- Droplet model (DM) [17], which is the basis of the Howard-Möller fission-barrier calculations [12],
- Finite-range liquid drop model (FRLDM) [13,20],
- Thomas-Fermi model (TF) [14,18],
- Extended Thomas-Fermi model (ETF) [15].

In order to test the consistency of these models, we study the difference between the experimental saddle-point mass ($E_f^{\text{exp}} + M^{\text{exp}}$) and the macroscopic part of the saddle-point mass ($E_f^{macro} + M^{macro}$), with $E_f$ being the height of fission barrier and $M$ the ground-state mass (see Fig. 2 for the definitions of the different variables):

$$\delta U_{sad} = E_f^{\text{exp}} + M^{\text{exp}} - (E_f^{macro} + M^{macro}). \quad (1)$$

Eq. (1) represents the most direct test of the model, as it does not refer to empirical or calculated ground-state shell corrections. According to the above-mentioned topographical arguments of Myers and Swiatecki [18] for nuclei with non-vanishing macroscopic fission barriers the experimental saddle-point masses should be very close to the values calculated by the macroscopic theory, i.e. the saddle-point shell effects – equivalent to $\delta U_{sad}$, Eq.(1) – should be very small, much smaller than ground-state shell effects. Of course, shell effects will change the deformation corresponding to the saddle point, but we are here interested in the mass at the saddle-point and not in its position in the potential-energy landscape. Moreover, as shell effects have a local character, $\delta U_{sad}$ should show only local variations and a mean value close to zero when followed over a wide range of neutron number. Any global trend should be included in the macroscopic model. Therefore, a general trend in $\delta U_{sad}$ with respect to the neutron content resulting from our analysis would indicate severe shortcomings of the model in extrapolating to nuclei far from stability.

---

[1] The macroscopic part of the fission barriers given by Howard and Möller is based on the droplet model [17].

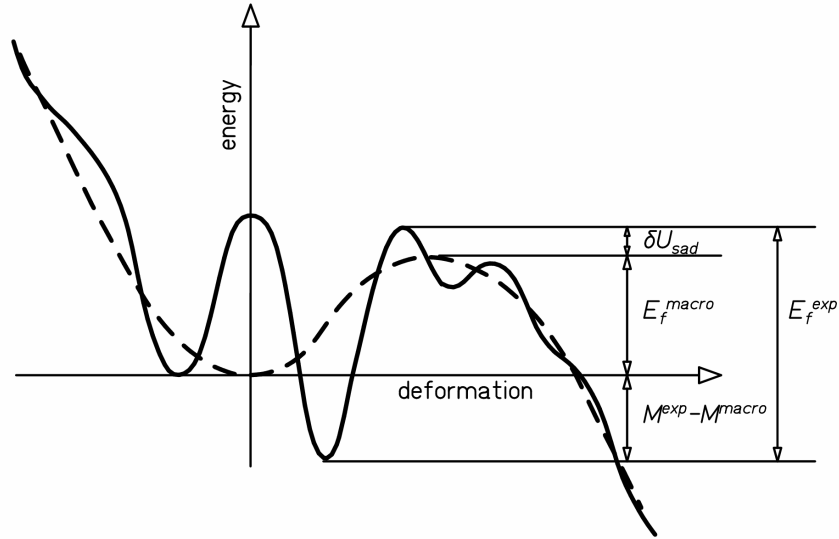

**FIGURE 2.** Schematic diagram of macroscopic (dashed line) and macroscopic-microscopic (full line) energy with definitions of several variables used in Eq. 1.

Fig. 3 shows a survey of the nuclei used for the present study on a chart of the nuclides. The experimental ground-state masses result from the Audi and Wapstra 2003 compilation [21], while the experimental fission barriers for these nuclei are taken from the compilation of Dahlinger *et al.* [19]. We have taken into account only the highest experimental fission barrier for two reasons: firstly, this barrier is determined experimentally with much less ambiguities than the lower barrier and, secondly, according to the topographic theorem [18], it should also be closer to the macroscopic barrier. Due to the large uncertainties in the measured fission barriers of lighter nuclei, we have considered only the nuclei with atomic number $Z \geq 90$. The wide span of the available data over neutron number, see Fig. 3, guarantees the global aspect of the study.

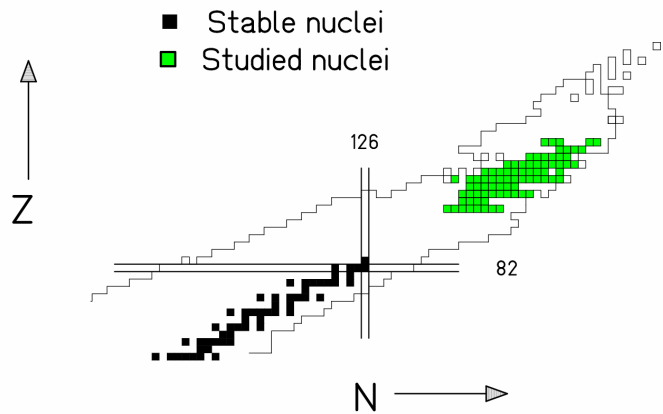

**FIGURE 3.** The nuclei – represented by grey squares – that were studied in the present work.

## RESULTS AND DISCUSSION

For the case of uranium isotopes, the variable $\delta U_{sad}$ as defined by Eq. (1) is shown in Fig. 4 as a function of the neutron number $N$. The FRLDM and the Thomas-Fermi model result in a quite similar behaviour of $\delta U_{sad}(N)$ with slopes close to zero. On the contrary, the results from the droplet model (DM) show that $\delta U_{sad}$ increases strongly with the neutron number, while the ETF model predicts a decrease. For this analysis, we did not have the macroscopic ETF ground-state masses available. Therefore, we have used the Thomas-Fermi masses from Ref. [18]. This is justified by the fact that the macroscopic part of the ground-state masses in the different models is adjusted to the large body of existing data, and different models predict very similar values and tendencies as a function of neutron number for the macroscopic ground-state masses (at least in the region of masses where the present analysis is performed). This was checked by comparing the results from the FRLDM, the DM and the TF model.

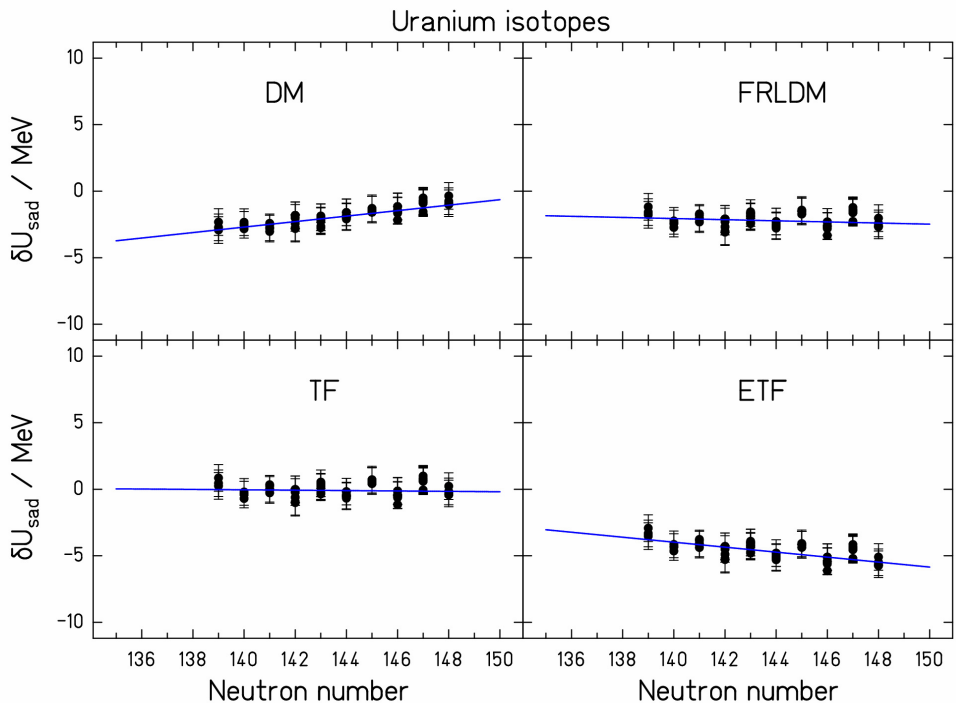

**FIGURE 4.** Difference between the experimental and the macroscopic part of the saddle-point mass calculated with the droplet model, the finite-range liquid-drop model, the Thomas-Fermi model and the extended Thomas-Fermi model for different uranium isotopes. The lines represent linear fits to the data.

If we would extrapolate the behaviour of $\delta U_{sad}$ from Fig. 4 to the case of e.g. $^{300}$U, which could be encountered on the r-process path [5], in case of the ETF model one would get an increase in the macroscopic barrier relative to $^{238}$U of ~ 8 MeV. This value is obtained, as mentioned above, when combining the fission barriers from the ETF model and the ground-state masses from the TF model. If we combine the ETF barriers with the ground-state masses from the FRLDM or the DM, this change in the barrier height from $^{238}$U to $^{300}$U amounts to ~10 or ~6 MeV, respectively, showing that the choice of the macroscopic ground-state mass model only plays a minor role in our analysis compared to the fission-barrier model. In case of the DM, for $^{300}$U one would obtain a decrease of ~ 10 MeV in the macroscopic barrier, leading, in fact, to no macroscopic barrier for this nucleus.

We applied the same procedure for all nuclei indicated in Fig. 3. The extracted slopes ($A_1$) of $\delta U_{sad}$ as function of the neutron excess are shown in Fig. 5 for the different elements. A similar behaviour of $\delta U_{sad}(N)$ as seen for uranium is also seen for the other elements. For all studied nuclei the droplet model predicts an increase in $\delta U_{sad}$ as a function of neutron excess, and, thus, positive values of $A_1$. This would imply that the macroscopic fission barriers decrease too strongly with increasing neutron number for all studied elements. The value of the mean slope averaged over the studied $Z$ range is 0.16 ± 0.01 MeV, indicating that the description of the isospin dependence of saddle-point masses within the droplet model is not consistent. The same conclusion was obtained in Refs. [19,22]. This result sheds a doubt on the applicability of the Howard-Möller tables of fission barriers [12] in regions far from stability. This finding is consistent with the analysis of the abundances produced in nuclear explosions performed by Hoff in 1987 [23], which also gave evidence that the Howard-Möller fission barriers of neutron-rich nuclei are too low. In case of the ETF model we had available the macroscopic barriers for the uranium isotopes only. Already in this case we see, Fig. 4, that over the range of 10 studied isotopes a clear correlation between $\delta U_{sad}$ and neutron number exists. For other elements, the values of the slopes in Fig. 5 were obtained by using the difference between the slope values for $Z = 92$ from the ETF model and the TF model, and by subtracting this difference from the TF slope values for other elements. We think that this procedure is well justified since the values of the slopes obtained with the DM, the FRLDM and the TF model show the same shape of the dependence on $Z$ – for all studied elements the difference between any two models is almost independent of $Z$, see Fig. 5. For all studied elements the obtained slopes from ETF are negative, with the average value of -0.18 ± 0.02 MeV. Opposite to the DM, these negative values of $A_1$ point to too strong increase in the macroscopic fission barriers with neutron number, also indicating possible problems in the consistency of the ETF model in describing saddle-point masses. The FRLDM results in much smaller slopes of $\delta U_{sad}$ as compared to the DM and ETF models, with the average value of -0.03 ± 0.01 MeV. The smallest values of the slopes are given by the TF model, with the average value of 0.003 ± 0.010 MeV, thus, making the TF model to be preferred for the description of the fission barriers of exotic nuclei.

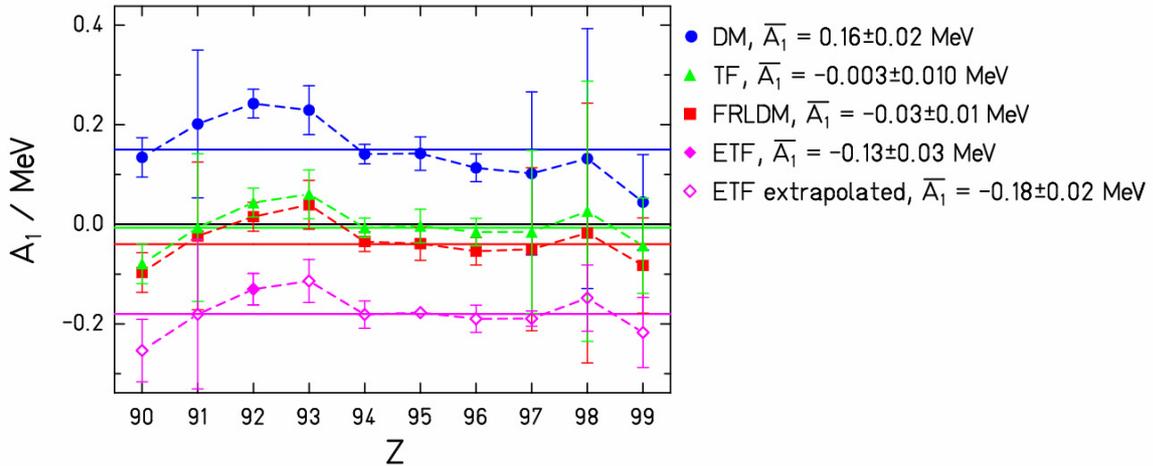

**FIGURE 5.** Slopes of $\delta U_{sad}$ as a function of the neutron excess are shown as a function of the nuclear charge number $Z$ obtained for the droplet model (points), the Thomas-Fermi model (triangles), FRLDM (squares) and the extended Thomas-Fermi model (full rhomboid); open rhomboids represent the values extrapolated from $Z=92$ in case of ETF, see text for more details. The full lines indicate the average values of the slopes. The average values are also given in the figure. Error bars originate mostly from the experimental uncertainties in the fission-barrier heights. Dashed lines are to guide eye.

One can, of course, raise the question of the origin of these differences between different models. Any of the mass models has a certain set (depending on the physical assumptions) of parameters – macroscopic and microscopic – whose values are obtained through a least-squares adjustment to the experimental ground-state masses and, in some cases, to the fission barriers. As the ground-state masses are influenced by strong shell effects, the values of the macroscopic parameters cannot be determined completely independently from the way how the microscopic part of the model is described. Moreover, in the least-squares adjustment, due to the much larger number of available experimental ground-state masses as compared to the small number of measured fission barrier heights, the weight of the former in the determination of the model parameters is much larger, and, consequently, the model parameters are mostly determined by the ground-state masses. This kind of the parameter adjustment can cause problems in calculating the fission barrier, as the barriers can be much more sensitive to specific parameters (e.g. curvature coefficient) than masses [15]. For example, the DM and the TF model use for the curvature coefficient $a_{cv}$ values of 0 MeV [17] and 12.1 MeV [18], respectively, which could be one source for the different predictions of these two models. Another parameter that has decisive influence on the fission-barrier heights is the symmetry-energy coefficient $J$. Mamdouh et al [15] discuss that if they use $J = 32$ MeV instead of their default value of 27 MeV they obtain a decrease in the height of the ETFSI fission barrier for $^{276}$U of 6 MeV, while for $^{238}$U the barrier is, in this case, decreased by only 1 MeV. This problem can be overcome, if the method proposed in the present work is applied, where the differences between experimental and calculated saddle-point masses would be used to fix the model parameters, thus avoiding any influence of ground-state shell corrections, which are difficult to model with the necessary precision. We expect that this powerful method will give much better constraints on a realistic description of the saddle-point masses and fission barriers for very neutron-rich nuclei.

## CONCLUSIONS

We have studied four different macroscopic models in order to check, whether they are adapted for predicting realistic values for the saddle-point masses of nuclei far from stability, in particular in model calculations of the r-process nucleosynthesis. The results of this study show that the most realistic predictions are expected from the Thomas-Fermi model [14]. The finite-range liquid-drop model [13] could still be applicable while inconsistencies in the saddle-point mass predictions of the droplet model [17] and the ETF model [15] were seen. This result raises severe doubts on the applicability of the Howard-Möllers fission-barrier tables [12] and the predictions of the ETFSI model [15] in modelling the r-process nucleosynthesis.

## ACKNOWLEDMENTS

We would like to thank Prof. Karlheinz Langanke for very fruitful discussions during the course of this work.